\documentstyle[11pt,aaspp4,flushrt]{aastex}

\slugcomment{\it Accepted for publication in the June 2010 Astronomical Journal}
\shortauthors{Eisenhardt et al.}
\shorttitle{$4.5\mu$m-Selected Ultracool Brown Dwarfs}

\def\tseven{SDWFS\,J142831.46+354923.1}
\def\tsevenshort{SDWFS1428+35}
\def\teight{SDWFS\,J143524.44+335334.6}
\def\teightshort{SDWFS1435+33}
\def\teightnew{SDWFS\,J143222.82+323746.5}
\def\teightshortnew{SDWFS1432+32}
\def\ydwarf{SDWFS\,J143356.62+351849.2}
\def\yd{SDWFS1433+35}
\def\spose#1{\hbox to 0pt{#1\hss}}
\def\simlt{\mathrel{\spose{\lower 3pt\hbox{$\mathchar"218$}}
     \raise 2.0pt\hbox{$\mathchar"13C$}}}
\def\simgt{\mathrel{\spose{\lower 3pt\hbox{$\mathchar"218$}}
     \raise 2.0pt\hbox{$\mathchar"13E$}}}

\def\plotfiddle#1#2#3#4#5#6#7{\centering \leavevmode
\vbox to#2{\rule{0pt}{#2}}
\includegraphics{#1}}

\begin{document}

\title{Ultracool Field Brown Dwarf Candidates Selected at ${\mathbf 4.5 \mu}$m}

\author{
Peter~R.~M.~Eisenhardt\altaffilmark{1},
Roger~L.~Griffith\altaffilmark{1},
Daniel Stern\altaffilmark{1},
Edward L.~Wright\altaffilmark{2},
Matthew~L.~N.~Ashby\altaffilmark{3}, 
Mark Brodwin\altaffilmark{3,4},
Michael~J.~I.~Brown\altaffilmark{5},
R.~S.~Bussmann\altaffilmark{6},
Arjun~Dey\altaffilmark{7},
A.~M.~Ghez\altaffilmark{2},
Eilat Glikman\altaffilmark{8}, 
Anthony~H.~Gonzalez\altaffilmark{9},
J.~Davy Kirkpatrick\altaffilmark{10}, 
Quinn Konopacky\altaffilmark{2},
Amy Mainzer\altaffilmark{1}, 
David~Vollbach\altaffilmark{9},
Shelley A.~Wright\altaffilmark{11}
}

\altaffiltext{1}{Jet Propulsion Laboratory, California Institute
of Technology, MS 169-327, 4800 Oak Grove Drive, Pasadena, CA 91109
[e-mail: {\tt Peter.Eisenhardt@jpl.nasa.gov}]}

\altaffiltext{2}{University of California, Los Angeles, CA 90095}

\altaffiltext{3}{Harvard-Smithsonian Center for Astrophysics, 60
Garden Street, Cambridge, MA 02138}

\altaffiltext{4}{W.~M.~Keck Postdoctoral Fellow at the Harvard-Smithsonian Center for Astrophysics}

\altaffiltext{5}{School of Physics, Monash University, Clayton, 
Victoria 3800, Australia}

\altaffiltext{6}{Steward Observatory, University of Arizona, Tucson AZ 85721}

\altaffiltext{7}{National Optical Astronomy Observatory, Tucson, AZ 85726}

\altaffiltext{8}{Divsion of Physics, Math and Astronomy, California
Institute of Technology, Pasadena, CA 91125}

\altaffiltext{9}{Department of Astronomy, University of Florida,
Gainesville, FL 32611}

\altaffiltext{10}{Infrared Processing and Analysis Center, California
Institute of Technology, Pasadena, CA 91125}

\altaffiltext{11}{Astronomy Department, University of California, Berkeley, CA 94709}

\begin{abstract}

We have identified a sample of cool field brown dwarf candidates using IRAC
data from the {\it Spitzer} Deep, Wide-Field Survey (SDWFS).  The
candidates were selected from 400,000 SDWFS sources with $[4.5] \leq 18.5$ mag
and required to have $[3.6]-[4.5] \geq 1.5$ and $[4.5] - [8.0] \leq 2.0$ 
on the Vega system.  The first color
requirement selects objects redder than all but a handful of presently
known brown dwarfs with spectral classes later than T7, while the
second eliminates 14 probable reddened AGN.  
Optical detection of 4 of the remaining 18 sources implies 
they are likely also AGN, leaving 14 brown dwarf candidates. 
For two of the brightest candidates (\teight\ and \teightnew), the spectral energy distributions
including near-infrared detections suggest a spectral class of $\sim$ T8.
The proper motion is $< 0\farcs25$ yr$^{-1}$, consistent with
expectations for a luminosity inferred distance of $>70$~pc.  The
reddest brown dwarf candidate (\ydwarf) has $[3.6] - [4.5]=2.24$ 
and $H - [4.5] > 5.7$,
redder than any published brown dwarf in these colors, and may be the first
example of the elusive Y-dwarf spectral class.  
Models from \citet{Burrows:03} predict
larger numbers of cool brown dwarfs should be found for a \citet{Chabrier:03} mass function.  
Suppressing the model [4.5] flux by a factor of two, as indicated by previous work,  
brings the Burrows models and observations into reasonable agreement.  
The recently launched Wide-field Infrared Survey Explorer ({\it WISE}) will probe a volume 
$\sim40\times$ larger and should find hundreds of brown dwarfs cooler than T7.

\end{abstract}


\section{Introduction
\label{sec:intro}}

Although first predicted to exist in 1963 \citep{Kumar:63, Hayashi:63},
brown dwarfs were not discovered until decades later.  The first
viable brown dwarf candidate was GD 165B \citep{Becklin:88}, an L
dwarf whose exact nature as hydrogen-burning star or brown dwarf
has yet to be ascertained \citep{Kirkpatrick:99}.  The first
undisputed brown dwarf, and the first T dwarf, was Gl 229B
\citep{Nakajima:95}, whose telltale methane absorption implied an
effective temperature too low for a normal star.  In the late 1990's
the advent of large-area surveys with near-infrared capability --
the Two Micron All-Sky Survey \citep[2MASS;][]{Skrutskie:06}, the
Sloan Digital Sky Survey \citep[SDSS;][]{York:00}, and the Deep
Near-Infrared Survey of the Southern Sky \citep[DENIS;][]{Epchtein:97}
-- uncovered hundreds more examples and enabled the study of brown
dwarfs as a population in their own right \citep{Kirkpatrick:05}\footnote{see {\tt
DwarfArchives.org} for a full list}.

The latest spectral type brown dwarfs currently known are T8 and T9 dwarfs 
found by 2MASS, the United Kingdom Infrared Deep Sky Survey 
\citep[UKIDSS;][]{Lawrence:07}, 
and the Canada France Brown Dwarf Survey \citep{Delorme:08b}. 
The coolest of these have effective temperatures of $\sim$550 K 
and implied masses of around 15-35 $M_{\rm Jupiter}$ 
for assumed ages of 1-5 Gyr 
\citep{Warren:07, Burgasser:08, Burningham:08, Delorme:08a, Leggett:09}.
Cooler field brown dwarfs must exist, however, as objects of 
much lower implied mass have been identified in young clusters 
such as the Orion Nebula Cluster \citep{Zapatero:02, Weights:08},
Upper Scorpius \citep{Lodieu:07} and Chamaeleon I \citep{Luhman:05}, 
or as companions to other low-mass cluster members 
\citep[e.g.,][]{Luhman:06}\footnote{see also {\tt http://vlmbinaries.org}}.
Finding and characterizing such colder field objects 
will set important boundary conditions on star formation processes  
and determine the total amount of mass in stars, 
a key ingredient in modeling galaxy formation.  
Identifying examples of such objects is also important 
to the study of very cold, planet-like atmospheres. 
A leading question is whether a new spectral class beyond T, 
dubbed ``Y'', will be needed \citep{Kirkpatrick:08}.

The two shortest wavelength bands in the {\it Spitzer} Infrared
Array Camera \citep[IRAC;][]{Fazio:04a} were designed to identify
cool brown dwarfs from the signature due to strong methane absorption
at $3.6 \mu$m coupled with a relative lack of absorption at $4.5
\mu$m \citep{Fazio:98}.  Finding the coolest and nearest brown
dwarfs is a key objective for the {\it Wide-field Infrared Survey
Explorer} \citep[WISE,][]{Liu:08}, which launched 2009 Dec. 14,
and hence two of its four imaging bands are at similar wavelengths
(3.4 and $4.6 \mu$m).  While a number of brown dwarf companions
have been found using IRAC, prior to the {\it Spitzer} Deep,
Wide-Field Survey \citep[SDWFS;][]{Ashby:09} only a single isolated
field brown dwarf, of spectral class T4.5, has been identified to
date on the basis of IRAC data \citep{Stern:07}.  This object,
IRAC~J1429050.8+333011, was found in the IRAC Shallow Survey
\citep{Eisenhardt:04} and was required to be unresolved in complementary
NOAO Deep Wide-Field Survey \citep[NDWFS;][]{Jannuzi:99} $I$-band
data, which necessitated $I < 23$.  Here we remove the limitation of 
requiring optical detection and use the deeper and more reliable
SDWFS IRAC data to search for cooler brown dwarf candidates.  

\section{Data and Selection Criteria}

\subsection{SDWFS\label{sec:select}}

SDWFS is a four epoch Legacy survey of 10 square
degrees in Bo\"otes using the IRAC instrument.  Each epoch covers
the entire field with three exposures separated by hours, each 30s
long, providing 12 observations at each sky location in all four
IRAC bands.  The first epoch is the IRAC Shallow Survey
\citep{Eisenhardt:04} from January 2004, and the last was obtained
in March 2008.  The publicly released\footnote{\tt
http://ssc.spitzer.caltech.edu/spitzermission/observingprograms/legacy/sdwfs}, 
full-depth (i.e., four epoch) catalogs contain 8.2, 6.7, 3.1, and 1.8 $\times 10^5$ 
distinct sources detected at 3.6, 4.5, 5.8, and 8.0~$\mu$m, of which
6.70, 5.28, 1.34, and 0.92 $\times 10^5$ exceed the average $5\sigma$,
aperture-corrected limits of 20.0, 19.0, 16.7, and 15.9 Vega mag.  
The uncertainties properly account for errors due to  
correlated pixels that arise during coadding.
See \citet{Ashby:09} for details of the SDWFS observations and analysis. 
We use the notation [3.6], [4.5], [5.8], and [8] for the Vega 
magnitudes in the four IRAC bands.

Since the Bo\"otes field is at Galactic latitude 67 degrees, the bulk of the 
[3.6] and [4.5] sources at these fluxes are extragalactic.   Figure 13 of
\citet{Ashby:09} shows nearly all of the sources have $-0.1 < [3.6] -[4.5] < 1$,
a range which includes the expected colors of galaxies out $z > 3$.  
AGN can extend to somewhat redder $[3.6] -[4.5]$ colors, 
while occupying a narrow range in $[5.8]-[8.0]$ \citep[Fig. 1 of][]{Stern:05b}.

Cool brown dwarf candidates were identified from the 671,688 SDWFS
$4.5\mu$m sources using the following selection criteria: 
(i) $[4.5] \leq 18.5$ (419,980 sources remaining), 
(ii) $[3.6] - [4.5] \geq 1.5$ (2,364 sources remaining), 
(iii) coverage of $\geq 10 \times$ 30~s in a 4\farcs2 $\times$ 4\farcs2 
($5\times5$ resampled pixels) region around each source 
in both the 3.6 and 4.5~$\mu$m bands (52 sources remaining).  
Photometry was measured in 3\arcsec\ diameter
apertures, corrected to 12\arcsec\ radius total Vega magnitudes.
The [4.5] magnitude limit provides $\sim 65\%$ completeness \citep{Ashby:09}, 
and $\sim 0.3$ mag color accuracy for objects which satisfy the color
limits.  The second criterion was selected to avoid confusion with
AGN, which are rare at $[3.6] - [4.5] > 1.5$ \citep{Stern:05b}, and
should identify brown dwarfs later than approximately spectral type
T7 \citep{Patten:06}.  The coverage map requirement reduces spurious
sources selected near the edges of the survey field, or heavily
affected by cosmic rays.

The 52 candidates identified with these three criteria were visually
inspected using separate images and photometry available for each
of the four SDWFS epochs, and 20 were classified as artifacts due
to glints, cosmic rays, diffraction spikes, or muxbleed trails from
bright stars.  Although all of the 52 candidates were observed in
epoch one (the IRAC Shallow Survey), with only three exposures there
are many more spurious candidates at these extreme colors, making
it impractical and ambiguous to visually screen them.  The T4.5 brown dwarf found
by \citet{Stern:07} used independent optical NDWFS data to ensure
reliability, but for cooler brown dwarfs, optical detection is not
expected.  The additional SDWFS exposures enable reliable detection
using IRAC data alone.

Many of the remaining 32 sources were suspiciously prominent at
$5.8\mu$m and $8\mu$m, leading to the imposition of an additional
criterion: (iv) $[4.5] - [8.0]  \leq 2.0$.  This final criterion is
designed to exclude heavily reddened AGN or dust obscured galaxies
\citep[DOGs;][]{Dey:08} as well as AGB stars which meet the second
criterion, but continue to brighten in the longer wavelength
IRAC passbands, unlike brown dwarfs (Figure~\ref{fig:colcol}).
Left-pointing arrows in Figure~\ref{fig:colcol} are based on two sigma 
upper limits at [8.0] based on the SDWFS depths in \citet{Ashby:09}.

This selection leaves the 18 candidates shown in Table~1.  Two are
noted as less robust based on visual inspection, and four show evidence of
being DOG variants (\S2.2), and hence are grouped separately at the
bottom of Table~1.  SDWFS images are available via the link in the footnote
provided earlier in this section.  Figure~\ref{fig.stamps}
shows the brightest and reddest candidates, as well as the reddest DOG
(SDWFS~J143819.58+340957.3, the only DOG with $[3.6] - [4.5] > 2$).

\subsection{Data at Other Wavelengths
\label{sec:OtherData}}

Most of the candidates have NDWFS photometry available in $B_W$,
$R$, and $I$, from the NDWFS, and $JHK_s$ photometry from the NEWFIRM survey
of the field with the KPNO 4m
(Gonzalez et al., in prep.), as shown in Table~1. The depth of these 
groundbased data are not as uniform as the SDWFS IRAC data,  
and hence optical and near-IR limits were estimated for each undetected
source.  When the estimated error exceeded 0.5 mag, two sigma 
upper limits above the measured flux (or 0 if measured flux was negative) 
were calculated from the errors in 3\arcsec\ diameter apertures, 
corrected to total magnitudes, for the appropriate location in the 
NDWFS and NEWFIRM data. 
Four of the candidates have faint or marginal detections
in $B_W$ (one in $R$ as well), and some of those 
have hints of detections in IRAC 5.8 or $8.0\mu$m.  Optical
detections giving $B_W - [4.5] \sim 7$ to 8 are not expected even for warm 
brown dwarfs, so it is likely these are variants of DOGs, 
which tend to be $z\sim2$ galaxies
and are detectable in $B_W$ from their Ly-$\alpha$ emission.  Indeed,
of the 14 objects classified as DOGs because they passed the first
three selection criteria and the visual inspection, but failed the
fourth, 10 are detected in $B_W$.  Hence we have separated the four objects
with faint optical detections in Table~1 from the other SDWFS brown dwarf candidates
with the heading ``Likely AGN." 

All but one of the 18 objects in Table~1 has {\it Spitzer} MIPS
$24\mu$m photometry available \citep{Houck:05},  
and none were detected to a level of 0.3 mJy.   
In contrast, of the 13 DOG candidates observed at $24\mu$m, 
10 were detected.

Likewise, none of the brown dwarf candidates were detected by the
{\it Chandra X-Ray Observatory} survey of the Bo\"otes field
\citep[XBo\"otes;][]{Kenter:05} or at radio wavelengths by either
the Faint Images of the Radio Sky at Twenty centimeters survey
\citep[FIRST;][]{Becker:95} or by the deeper Westerbork 1.4~GHz
observations of 7~deg$^2$ of Bo\"otes reported in \citet{deVries:02}.
Four DOG candidates were detected at radio wavelengths,
one of which was also detected by XBo\"otes.
That MIPS-selected source, SDWFS~J143644.23+350627.0,  
has a redshift of $z=1.95$ from {\it Spitzer} Infrared
Spectrograph observations \citep{Houck:05}.

\subsection{Near-IR Follow-Up}

The brightest two candidates (see Figure~\ref{fig.stamps})
were targeted for additional follow-up using the Wide-field Infrared
Camera \citep[WIRC;][]{Wilson:03} at the Palomar 5.08~m telescope on UT 2008 Aug 25
(\teight, hereafter \teightshort) and UT 2008 Aug 28 (\tseven,
hereafter \tsevenshort).  Dithered sets of $4\times30$s images were
taken with exposure times (seeing) of 36m (1\farcs1) at $J$ and 54m
(1\farcs0 - 1\farcs2) at $H$ for \tsevenshort, and 54m (1\farcs3) at $H$
for \teightshort.  Photometry was calibrated using $\sim 10$ 2MASS sources
in each field.  No significant detections were obtained.
Using the rms variation in 3\arcsec\ diameter
apertures, the $2\sigma$ aperture corrected limits are $J>21.9$ and
$H>21.4$ for \tsevenshort.  For \teightshort\ the Palomar data yield $H>21.3$, 
but the NEWFIRM survey provides 
detections at $J=21.16 \pm 0.13$ and $H=21.09 \pm 0.48$ as listed in Table~1.  
The NEWFIRM survey also detected \teightnew\ at $J=21.17 \pm 0.18$. 
Both the \teightshort\ and \teightshortnew\ detections are at levels expected
for ultracool brown dwarfs, and all of the objects in Table~1 have near-IR to [4.5] colors
or limits consistent with late T dwarfs (\S \ref{sec:BDs}).


The reddest candidate identified (\ydwarf, hereafter \yd) 
was targeted for followup with the NIRC2 camera (P.I. K. Matthews) on the Keck~II telescope 
using laser guide star adaptive optics \citep{Wizinowich:06, vanDam:06} on UT 2009 June 11.  
A total of 42 minutes of integration using 3 minute exposures
and a pixel scale of $0\farcs0397$ in $H$ was obtained under photometric 
conditions and $0\farcs5$ seeing.  An $R=17.6$ star 20\farcs7\
to the West was used to provide tip-tilt correction.  The point source FWHM 
in the combined image is $0\farcs12$, and the estimated Strehl ratio is 0.2
at the location of \yd.  The field of view was positioned so that the tip-tilt
star was in the field to provide a photometric and astrometric reference, but
it was slightly saturated in 3 minutes, so additional $4 \times 30$s coadded
exposures were obtained to calibrate the photometry.  No detection of \yd\
is apparent in the combined image (Figure~\ref{fig:y-dwarf}).  
We estimate $H > 24.2$ for \yd, by comparing to scaled down versions of 
the photometric tip-tilt star (which has $H=16.19$ from 2MASS) added into the combined image. 

Figure \ref{fig:BD_sed} shows the SEDs for \teightshort\ 
and \yd.  

\subsection{Proper Motions\label{sec:prop_mot}}

With low intrinsic luminosity, cool brown dwarfs should be nearby
and thus may have detectable proper motions in the four years spanned
by SDWFS.  Sources with large proper motion might even be rejected
from the SDWFS catalog because they move between epochs.  To allow
for this, a search was made for objects in each of the four SDWFS
epochs which satisfied criteria (i) and (ii), and whose positions
matched to within 10\arcsec.  This search did not find any sources
not already identified using the full SDWFS dataset as described
above.

The average astrometric frame offset between SDWFS epochs is
$\approx 0\farcs17$, with a standard deviation of $\approx 0\farcs35$
for sources with $[4.5] < 18$ \citep{Ashby:09}.  
For sources near the $[4.5] = 18.5$ limit of the present sample,
a standard deviation of $\approx 0\farcs55$ is appropriate.
None of the brown dwarf candidates in Table~1
show significant ($\geq 3\sigma$) proper motions, implying proper
motions $\mu \leq 0\farcs25\, {\rm yr}^{-1} (0\farcs4\, {\rm yr}^{-1}$
for sources near the $[4.5] = 18.5$ limit).  

\citet{Ashby:09} do find four SDWFS sources with proper motions of $\approx 0\farcs3\,
{\rm yr}^{-1}$, including two\footnote{
SDWFS~J142723+330403 was mistakenly identified with a nearby but unrelated 
NDWFS source in \citet{Ashby:09}, and the $B_WRI$ photometry shown for this 
object in Table\,26 of that paper is incorrect.  Instead, only the NDWFS 
upper limits apply.} 
with $[3.6] - [4.5]$ colors appropriate for mid- to late-T brown dwarfs \citep{Patten:06}, but
not meeting the $[3.6]-[4.5] \geq 1.5$ criterion.
The NEWFIRM survey of the SDWFS field provides 
$J=19.48$ and $H=19.94$ in aperture corrected 4\arcsec\ diameter apertures for SDWFS~J142723+330403,
consistent with expectations for a T7 dwarf with $[4.5]=16.96$.

\section{Discussion}

The SDWFS search confirms the impression from the IRAC Shallow Survey 
\citep[][Fig. 4b]{Eisenhardt:04} that at high Galactic latitude, 
objects with IRAC colors as red as the coolest known brown dwarfs 
are rare.  Of 367,176 SDWFS sources meeting criteria (i) and (iii) in \S\ref{sec:select},
less than one in ten thousand is a real source meeting criterion (ii), i.e.  
($[3.6] - [4.5] \geq 1.5$, equivalent to $F_{\nu}(4.5)/F_{\nu}(3.6) > 2.5$).  
Only two real objects have $[3.6] - [4.5] > 2$ 
--- presumably the realm inhabited by the elusive Y-dwarfs --- 
making them an order of magnitude rarer still.
Although we were careful {\it not} to require [3.6] detection, all brown dwarf candidates, 
and all but one DOG (SDWFS~J143819.58+340957.3 --- Figure \ref{fig.stamps}) 
are in fact clearly detected in [3.6].

A blackbody with $[3.6] - [4.5] \geq 1.5$ would have $T_{\rm BB} \simlt 500$K, while
a power-law spectrum would need $\alpha > 4$ where $F_{\nu} \propto \nu^{-\alpha}$. 
Such spectra might arise from cool brown dwarfs or warm dust, 
or from obscuration of hotter spectra by dust.
For brevity, we often substitute $C$ for $[3.6] - [4.5]$ 
in the remainder of the discussion.



\subsection{Dusty Sources}

Dust enshrouded carbon stars and AGB stars \citep[e.g.][]{Cutri:89}, 
or class 0 or I protostars \citep[e.g.][]{Enoch:09}
can have very red IRAC colors due to warm dust emission (Figure \ref{fig:colcol}).  
However for the $T_{\rm BB} \simlt 500$K needed to produce
$[3.6] - [4.5] \geq 1.5$, the corresponding  $[4.5] - [8]$ is  $> 2.3$, 
violating criterion (iv). 
Significant emission from cooler dust is typical for such sources, which is even less
consistent with the longer wavelength photometry for the brown dwarf candidates.
Hence it is possible that some of the sources classified as DOGs may in fact be stellar. 

Nevertheless Figure~1 shows that the locus of likely AGB stars from \citet{Robitaille:07}
lies near the brown dwarf selection region, 
with three out of 23 with $C \equiv [3.6] - [4.5] \geq 1.5$ falling inside it.  
AGB stars associated with our Galaxy are unlikely contaminants, 
both due to their faintness and to the high Galactic latitude of the field (67 degrees).  
The absolute [4.5] magnitude of a typical AGB star is $\sim -10$ \citep{Cutri:89}, 
putting AGBs with the characteristic $[4.5] \sim 18$ values
found in Table~1 at a distance of order 3 Mpc.  
\citet{Mauron:08} finds three AGB stars more than 100 kpc from the Sun, 
presumably from the disruption of tidally captured dwarf galaxies.  
If this space density is typical, there could be of order 10 AGB stars 
in the SDWFS volume at a distance near 3 Mpc.  
However, using the NASA Extragalactic Database, 
we find no galaxies brighter than 15th mag (optical, i.e $\sim 1000\times$
less luminous than the Milky Way at 3 Mpc) 
and with redshifts $< 1000$ km/s 
within 5 degrees ($\sim300$ kpc at 3 Mpc) of the field. 
 
Note that AGB stars are often large amplitude variables.
\citet{Rejkuba:03} and \citet{Davidge:04} find that most AGB stars 
in NGC 5128 and M32 respectively are variable.
\citet{Rejkuba:03} give an average K band amplitude of 0.77 mag and a period of 395 days, 
similar to values found by \citet{Glass:95} for Galactic AGB stars. 
The threshold for the \citet{Robitaille:07} sources is a factor of two (0.75 mag).
The peak to peak variation in the [4.5] mags 
between the four SDWFS epochs (which span four years) 
exceeds the factor of two level for two of the sources in Table~1. 
For one of these,  SDWFS~J143712.48+334516.5, inspection of the data 
shows this is because its [4.5] brightness 
is spuriously high in one epoch due to a cosmic ray in the aperture, 
which was rejected in the combined 4-epoch measurement.  
For the other source, SDWFS~J143821.36+353523.3, although the 
peak to peak variation is 1.5 mag, the rms is 0.66 mag, 
which is only slightly more than a one sigma excess above 
the median variability at [4.5] for this magnitude \citep{Kozlowski:10}.
Because there is little evidence for excess variability in the Table~1 sources, 
and no obvious source for intergalactic AGB stars, 
we consider such stars unlikely to be a significant contaminant
for our brown dwarf candidate sample.
  
\citet{Dey:08} describe a sample of 2603 objects in the Bo\"otes field 
selected to have $R - [24] \ge 14$, which they interpret as 
dust obscured galaxies (DOGs).  
Of the 2491 DOGs from \citet{Dey:08} with SDWFS IRAC photometry, 
12 have $C  \ge 1.5$, but all of these have 
$[4.5] - [8] > 2$ so they do not appear in Table~1.    
DOGs with $C \ge 1.5$ (either from the \citet{Dey:08} $R - [24]$ selection, 
or classified as DOGs here from due to $[4.5] - [8] > 2$)
are plotted as open circles in Figure \ref{fig:colcol}. 
No attempt is made to distinguish AGN with $C > 1.5$ from DOGs, 
since DOGs include AGN. Objects plotted as dots in Figure 1 satisfy the 
\citet{Stern:05b} AGN criteria.  Of these four have $C > 1.5$,  
and with $[4.5] - [8] \sim 3$ they fall in the midst of the DOG colors. 

\citet{Dey:08} suggest the $24\mu$m emission arises from
warm dust heated by an AGN for the brighter sources, or from redshifted PAH
emission at $z\sim2$ for the fainter sources.  
The red IRAC colors for these fainter DOGs are likely due to obscuration of stellar light
by dust. We estimate $A_V \simgt 6$ to produce $C \ge 1.5$ at $z \sim 2$.
For the reddest DOG, SDWFS~J143819.58+340957.3 
(Figures \ref{fig.stamps} and \ref{fig:BD_sed}), $A_V$ well above 10 is indicated.  
As noted in \S\ref{sec:OtherData}, {\em none} of the objects in Table~1 are
detected at $24\mu$m, while of the 14 sources classified here as DOGs
from IRAC photometry (i.e $C \ge 1.5$ and $[4.5] - [8] > 2$), 
10 of the 13 observed at $24\mu$m were detected.   

The 14 IRAC-classified DOGs were also detected in $B_W$ in
9 cases, likely due to redshifted Ly-$\alpha$ emission, 
with 8 of the $B_W$ detected objects having detections in $R$ and $I$ as well.  
The bottom portion of Table~1 lists three brown dwarf candidates 
(based on their IRAC photometry) with faint optical detections, and a fourth with a marginal
optical detection, under the heading ``Likely AGN."  These have been marked with
shaded gray circles in Figure \ref{fig:colcol} to indicate that they are likely DOG variants.
The potential for additional DOG/AGN contamination of the sample is discussed in
\S \ref{sec:complete}.

\subsection{Brown Dwarfs\label{sec:BDs}}

Even for $T_{\rm eff} < 500$K, brown dwarf spectra are expected to have $[4.5] - [8] < 2$
due to molecular absorption features 
\citep[Figures \ref{fig:colcol} and \ref{fig:BD_sed},][]{Burrows:03}.
If the sources in Table~1 are brown dwarfs, 
they have spectral classes later than T6, 
based on their $[3.6] - [4.5]$ colors \citep[Figure 10,][]{Patten:06}.  
Only two of the 86 sources in \citet{Patten:06} have $C \equiv [3.6] - [4.5] \geq 1.5$: 
GJ 570D (T7.5) with a color of 1.68, and 2MASS 0415-0935 (T8.0)
with a color of 1.82.  At the time this paper was submitted, only three other brown
dwarfs had colors this red: 
ULAS0034 with a color of 1.81 and spectral class T8.5 \citep{Warren:07}; 
ULAS1335 \citep[$C=2.05$, T9;][]{Burningham:08}; 
and 2MASS 0939-24 
\citep[$C=2.10$, a possible T8 binary;][]{Burgasser:08}\footnote{
After submission, five others appeared in \citet{Leggett:10}:
2MA 0348-60, $C=1.53$, T7; 
ULAS 2321+13, $C=1.64$, T7.5;
ULAS 1238+09, $C=1.75$, T8.5;
2MA 1114-26, $C=1.78$, T7.5;
and Wolf 940B, $C=2.01$, T8.5.
}.
If it is confirmed, with $C = 2.24$ \yd\ would have the reddest IRAC color
of any brown dwarf yet found (Figure \ref{fig:colcol}).

While the optical photometry does not reach the depths expected 
for T  and late L dwarfs with
$[4.5] \simgt 18$, the typical $I - [4.5] \simgt 6$ limits argue against the sources in
Table~1 being significantly hotter than brown dwarfs. 
From Figure 8 of \citet{Hawley:02} and Fig. 10 of \citet{Patten:06},  
the onset of L dwarfs is at $i-[4.5] \sim 6.2$, equivalent to $I-[4.5] \sim 5.8$. 

The $J - [4.5]$ color ranges from 2.49 to 3.99 
for the previously published brown dwarfs with $C \geq 1.5$. 
For \teightshort\ we find $J - [4.5] = 3.56$, which together with 
$C = 1.84$ is very similar to the observed values for ULAS0034.  
Detailed modeling of ULAS0034 lead \citet{Warren:07} and \citet{Leggett:09} 
to conclude that it has $T_{\rm eff} \approx 600$K.  
For \teightshortnew\ the $J - [4.5] = 3.11$ and $C = 1.91$ are each about
0.1 mag redder than the corresponding 2MA 0415-0935 (T8.0) values.  
Figure \ref{fig:BD_sed} shows the photometry for \teightshort\ overplotted 
with a 600K model from \citet{Burrows:03}. 
The $J - [4.5]$ limits for other brown dwarf candidates range from $>2.9$ to $>4.4$,
consistent with spectral classes beyond T6.

\citet{Burningham:08}, \citet{Warren:07}, and \citet{Leggett:10} suggest that $H - [4.5]$
is tightly correlated with $T_{\rm eff}$.  
Figure~\ref{fig:hm4p5} shows $H - [4.5]$ as a function of spectral type.
The published brown dwarfs with $C \geq 1.5$ 
have $H - [4.5]$ colors ranging from 2.98 to 4.34, 
while the $2\sigma$ limits here range from $>2.1$ to $>3.7$ ($>5.7$ for \yd\, see below), 
consistent with spectral classes beyond T5.  
For \teightshort\ $H - [4.5] = 3.49$ is similar to the T8 brown dwarf 2MA 0415-0935. 
We include \teightshortnew\ in figure~\ref{fig:hm4p5} as a grey-shaded point with $H-[4.5]=3.46\pm0.7$.
This assumes $J-H \sim -0.35$ based on photometry 
for late-T objects from \citet{Patten:06} and \citet{Leggett:10} 
together with our $J=21.17$ detection to estimate $H=21.52$, 
with the error bar consistent with the observed $2\sigma$ limit of $H>20.8$.  

The available $K_s$ data offer fewer constraints than $J$ and $H$.
The published brown dwarfs with $C \geq 1.5$ 
have $K - [4.5]$ colors ranging from 2.98 to 5.17, 
while the $2\sigma$ limits here range from $K_s>1.1$ to $>2.4$, 
consistent with spectral classes beyond L5.  

From their $J - [4.5]$, $H - [4.5]$, and $[3.6] - [4.5]$ colors and limits, 
we associate a spectral class of T8 for \teightshort, 
and (more tentatively) T8.5 for \teightshortnew.  
Assuming $M_{4.5} = 13.5$ for \teightshort\ 
\citep[based on][]{Patten:06}, the luminosity distance is $\sim 70$ pc.  
As noted in \S \ref{sec:prop_mot}, neither source shows significant proper motion, 
with an estimated $0\farcs25\, {\rm yr}^{-1}$ limit for sources with $[4.5] < 18$. 
This is unsurprising, since typical proper motions should be 
$\sim 0\farcs1\, {\rm yr}^{-1}$ at this distance,  
using an average tangential velocity from a
volume-limited sample of stars of 37 km~s$^{-1}$ 
\citep[or $\sim 4$ AU in 6 months]{Reid:97}. 


With $C=2.24$, $[4.5] - [8] < 1.6$, and $H - [4.5] > 5.7$,
the SED for \yd\ falls in
previously unpopulated regions of color space 
(Figures \ref{fig:colcol} and \ref{fig:hm4p5}), 
and hence this object may be a member of the long-sought Y-dwarf class.
Figure \ref{fig:BD_sed} shows a 400K \citet{Burrows:03} model which
roughly agrees with the observations for \yd.
If the 1.3 mag drop in [4.5] luminosity between the 400K and 600K
\citet{Burrows:03} models is applicable, $M_{4.5} \sim 15$ for \yd\, 
implying a distance $\sim 50$ pc.    Again this is consistent with 
the $0\farcs4\, {\rm yr}^{-1}$ proper motion limit noted in \S \ref{sec:prop_mot}
for the $[4.5] = 18.47$ mag of \yd,
vs. the typical expected proper motion of $\sim 0\farcs15\, {\rm yr}^{-1}$ at 50 pc. 
Alternative explanations for \yd\ must account for the non-detections in [5.8], [8], and [24].

\subsection{Reliability and  Completeness\label{sec:complete}}

The presence of three (possibly four) optical detections amongst the 18 sources which
satisfy the IRAC color selection criteria shows that those criteria
do not produce a pure brown dwarf sample, and suggests that additional
AGN /DOGs may have scattered into the brown dwarf selection region.
Monte-Carlo simulations using the existing SDWFS catalog and error
distribution were carried out to evaluate the expected level of
such contamination.

The pool of 11,907 SDWFS sources was identified which satisfied
criterion (iii) in \S \ref{sec:select} and relaxed versions of
criteria (i) and (ii), i.e.  $[4.5] \leq 19.0$ and $C \equiv [3.6] - [4.5]
\geq 1.0$. For each source in this pool, the associated flux errors
was used to generate 10,000 realizations of the IRAC photometry and
to find the likelihood that each source would meet the full color
selection criteria given in \S \ref{sec:select}.  The summed
likelihood of was 67.3, with 34 sources having likelihoods greater
than $50\%$ (and a summed likelihood of 26.3). The summed likelihood
of selection for the 5202 sources with $1.0 < C < 1.1$ was less than 0.1, 
indicating contamination from bluer sources is not important.  
Note that $C > 1$ corresponds to brown dwarf spectral types 
later than T5 \citep{Patten:06, Leggett:10}.
Visual inspection was carried out in the same manner described in \S \ref{sec:select}
for the 86 sources with individual likelihood $\geq 20\%$, and for
40 representative sources with lower likelhooods.  Based on this
inspection, one third of the sources meeting the color criteria
would be classified as artifacts, with the remainder equally divided
between objects in or very near the cold brown dwarf color selection
space, objects with AGN colors scattering into the selection criteria
(e.g., with faint detections at $5.8$ and $8.0\, \mu$m), and objects
for which the distinction between AGN and cold brown dwarf colors
was uncertain. We infer from this that 1/3 to 2/3 of the 18 objects
in Table~1 are likely to be objects whose true colors are consistent
with cold brown dwarfs, (i.e. that 6 to 12 of the 14 sources in
the upper part of Table~1 are likely real brown dwarfs).

A complementary Monte-Carlo calculation was made to assess the
completeness of the color-selected sample given typical photometric
errors as a function of magnitude for the SDWFS data.  The probability
that a source would meet the color selection criteria was evaluated
using 10,000 realizations of sources as a function of magnitude and
color over the range $15 \leq [4.5] \leq 19.5$ and $1.0 \leq [3.6]
- [4.5] \leq 2.6$.  As expected, the probability is $\sim50\%$ for
bright sources with $[3.6] - [4.5] = 1.5$.  
For objects $[3.6] - [4.5] = 1.6$, 90\% will be selected at $[4.5] = 17.0$, and 90\% of
objects with $[3.6] - [4.5] = 1.9$ are selected at $[4.5] = 18.0$.
Applying the appropriate percentages as a function of magnitude
and color to the sources in the upper part of Table~1, we find an 
average completeness of $\sim60\%$.   

However it is also the case that warmer brown dwarfs whose
true color is bluer than $C=1.5$ can scatter into the sample.
Using models for the true distribution of brown dwarf magnitudes
and colors (\S \ref{sec:counts}) in conjunction with the completeness
calculations, we find that this effect closely compensates for losses
due to incompleteness.  With the finding 
that sources hotter than T6 do not contribute appreciably, 
and the reliability estimate, 
this implies that the true population of cool brown dwarfs
meeting the selection criteria is between 6 and 12. 
Some of these are likely to be unresolved binary brown dwarf systems 
\citep[see e.g.][]{Burgasser:07}, but this has a relatively small effect on
the number density because the increase in numbers due to
binaries is compensated for by the larger volume over which they
are detectable in a flux limited sample.   If a fraction $B$ of the sample
is equal mass binaries, the net effect is a reduction of 
$B(1 - \sqrt(2)/2) \approx 0.3B$ in a volume limited sample.
Hence we do not correct
for binarity, and in the following section we take 9 brown dwarfs 
with $C > 1.5$ as representative, of
which 8 have $1.5 < C < 2$ and one has $C > 2$.

\subsection{Brown Dwarf Counts\label{sec:counts}}

We compare our source counts to
the models of \citet{Burrows:03},
who give a grid of 32 cool brown dwarf models with cooling curves 
parameterized by mass $\mu$ and age $t$.  
From the tabulated effective temperature $T_{\rm eff}$, gravity $g$ and mass we
have computed the luminosity $L$, and set up a linear interpolation in
$\log L$ and $\log T_{\rm eff}$ {\it vs.} $\log \mu$ and $\log t$ to
give the luminosity and effective temperature for any mass and age.

The expected number $N$ of detectable brown dwarfs can be computed for
any mass function and age distribution using the distance $r$ as a function
of brown dwarf magnitude $m$  
\begin{eqnarray}
r & = & 10^{1+(m-M(\mu,t))/5} \nonumber \\
N & = & \Omega \int \int \int p[m,C(\mu,t)]\ n(\mu,t)\ r^2 \frac{dr}{dm}\ dm\ d\mu\ dt
\end{eqnarray}
where 
$\Omega$ is the survey area, 
$M(\mu,t)$ is the absolute magnitude as a
function of mass and age, 
$C(\mu,t)$ is the color as a function of mass and age, 
$n(\mu,t)$ is the number density of
brown dwarfs per unit mass and age,
and $p(m,C)$ is the probability 
from the Monte-Carlo completeness calculation in \S \ref{sec:complete}
of a brown dwarf with magnitude $m$ and color $C$ being selected. 
In general the number of sources
scattered into the acceptance region was quite similar to the number
scattered out of the region, so the Monte-Carlo completeness corrections
were small.  

Assuming a uniform distribution
in age between 100 million and 10 billion years, a \citet{Chabrier:03}
log normal mass function peaking at $0.079 M_\odot$, 
$\Omega = 10\, {\rm deg}^2$, and a magnitude limit of
$[4.5] < 18.5$ or a flux $> 7.15 \mu$Jy, the predicted numbers are
55 sources with $1.5 < C < 2$ (where $C \equiv [3.6]-[4.5]$) and 63
with $C > 2$.  Since we estimate only 8 sources are brown dwarfs 
with $1.5 < C < 2$ and one
with $C > 2$, the hypothesis that \citet{Chabrier:03} and
\citet{Burrows:03} are both correct can be rejected.  There
are two problems: the predicted ratio $N(C > 2)/N(1.5 < C < 2) \approx 1.1$ 
is much higher than the observed $1/8$, and the predicted
$N(1.5 < C < 2)$ is too high.  While power law mass functions ($n 
\propto M^{-\alpha}$) with $\alpha$ near 0 predict lower counts,
the ratio of counts in the color bins is still too high, and has
only a weak dependence on $\alpha$.

\begin{figure}
\plotfiddle{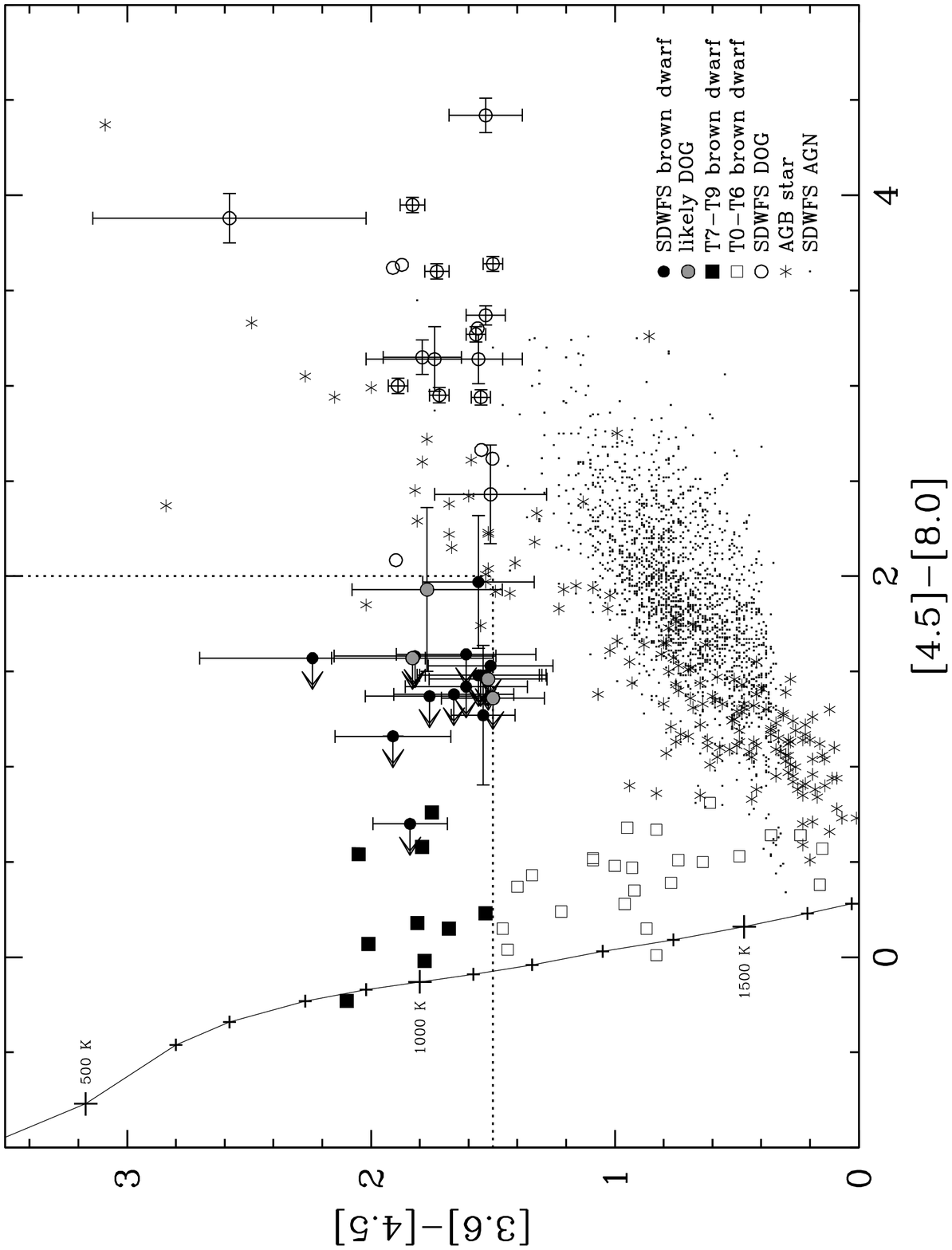}{5.0in}{-90}{70}{70}{-280}{420}
\vspace*{0cm}
\caption{IRAC color-color diagram illustrating the selection criteria
applied to identify the coolest brown dwarfs (dotted lines).  
The solid line shows predicted brown dwarf colors models from 
\citet{Burrows:03} for $T_{\rm eff} \leq 600$K 
using nonequilibrium, $\log g=5$ models, 
and from \citet{Hubeny:07} for $T_{\rm eff} > 600$K
using nonequilibrium, $\log g=5$, $\log K_{zz}=2$, 
``fast2" $\rm{CO/CH_4}$ reaction speed models,  with tick marks at 100K
intervals from 500 to 1700K. 
Dots show mid-IR selected AGN selected from the Bo\"otes field
using the \citet{Stern:05b} IRAC color criteria.  
Asterisks show variable $8\mu$m sources found near the Galactic plane by
\citet{Robitaille:07} and thought to be AGB stars. 
Open squares show spectral class T0 - T6 brown dwarfs from \citet{Patten:06}
while filled squares show cooler brown dwarfs from 
\citet{Patten:06}, \citet{Warren:07},
\citet{Burningham:08}, \citet{Burgasser:08}, and \citet{Leggett:10}.
Black circles show the 14 cool brown dwarf candidates identified in Table~1;
arrows show $2 \sigma$ upper limits based on SDWFS depth at [8]. 
Open circles show the 14 SDWFS sources with $[3.6] - [4.5] \geq 1.5$ and
$[4.5] - [8.0] \geq 2$, which are classified as DOGs.
Open circles without error bars are for several 
DOGs selected by \citet{Dey:08} using $R - [24] > 14$ 
which did not meet our SDWFS selection criteria (\S \ref{sec:select}).  
Gray circles are objects which meet the cool brown dwarf color criteria but 
for which optical detections suggest they are DOGs.  The brown dwarf candidate 
near the lower right edge of the selection criteria has a questionable [8] detection,
implying it likely has a bluer $[4.5] - [8]$ color than indicated.
\label{fig:colcol}}
\end{figure}

\begin{figure}
\plotfiddle{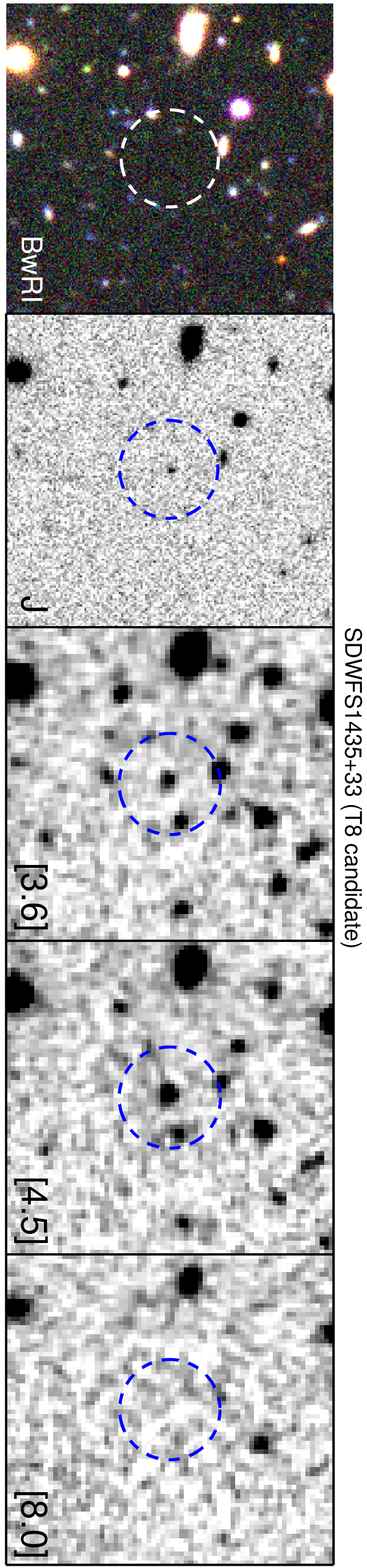}{5.5in}{90}{70}{70}{280}{90}
\plotfiddle{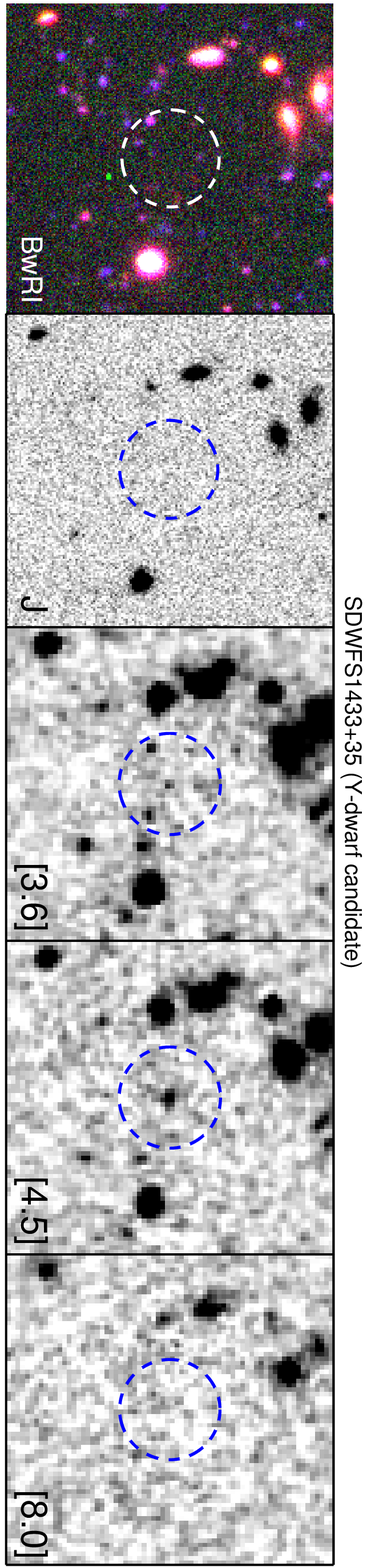}{0.0in}{90}{70}{70}{280}{-30}
\plotfiddle{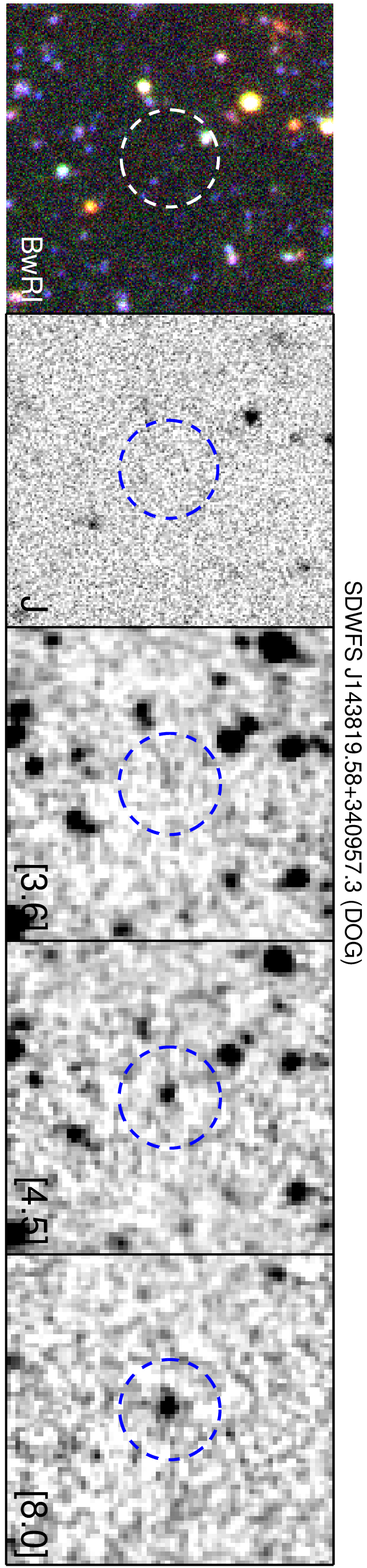}{0.0in}{90}{70}{70}{280}{-150}
\caption{Multi-wavelength images of sources with very red $[3.6] - [4.5]$
colors in the SDWFS field.  Images are 1 arcmin on a side, with
North up and East to the left.  The left panels are color composites
of the NDWFS $B_{\rm W}RI$ data, followed by $J$ images from the NEWFIRM survey
and IRAC data from SDWFS.  Circles are
10\arcsec\ in radius, centered on the red IRAC sources.  The top
strip shows the brightest cool brown dwarf (T8) candidate in
SDWFS, while the middle shows the reddest brown dwarf (Y-dwarf) candidate.  
The bottom strip shows the reddest DOG in SDWFS, 
which has $[3.6] - [4.5]=2.58$, but is much redder than the brown dwarf
candidates at longer IRAC wavelengths, with $[4.5] - [8.0]=3.88.$
\label{fig.stamps}}
\end{figure}

\begin{figure}
\plotone{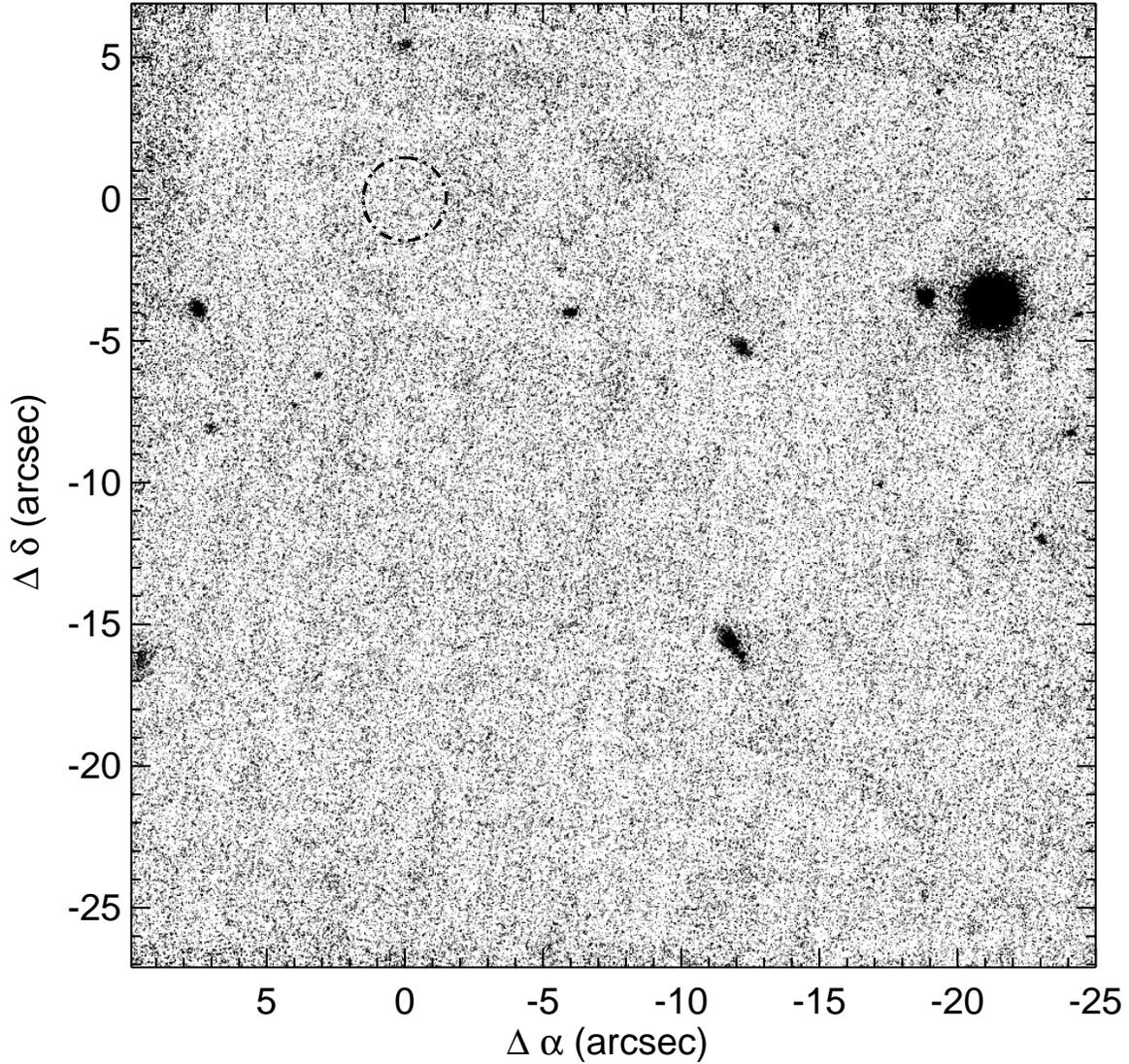}
\caption{$H$ image of the region near \yd\ (the Y-dwarf candidate) obtained with the 
laser guide star adaptive optics system and the NIRC2 camera on the Keck II telescope.  
The field is 34\arcsec\ on a side with North up and East left.  
The position of the [4.5] source is marked with a 3\arcsec\ diameter circle.
There are no sources with SNR $> 2$ within the circle, or with a plausible FWHM.  
\label{fig:y-dwarf}}
\end{figure}

\begin{figure}
\plotfiddle{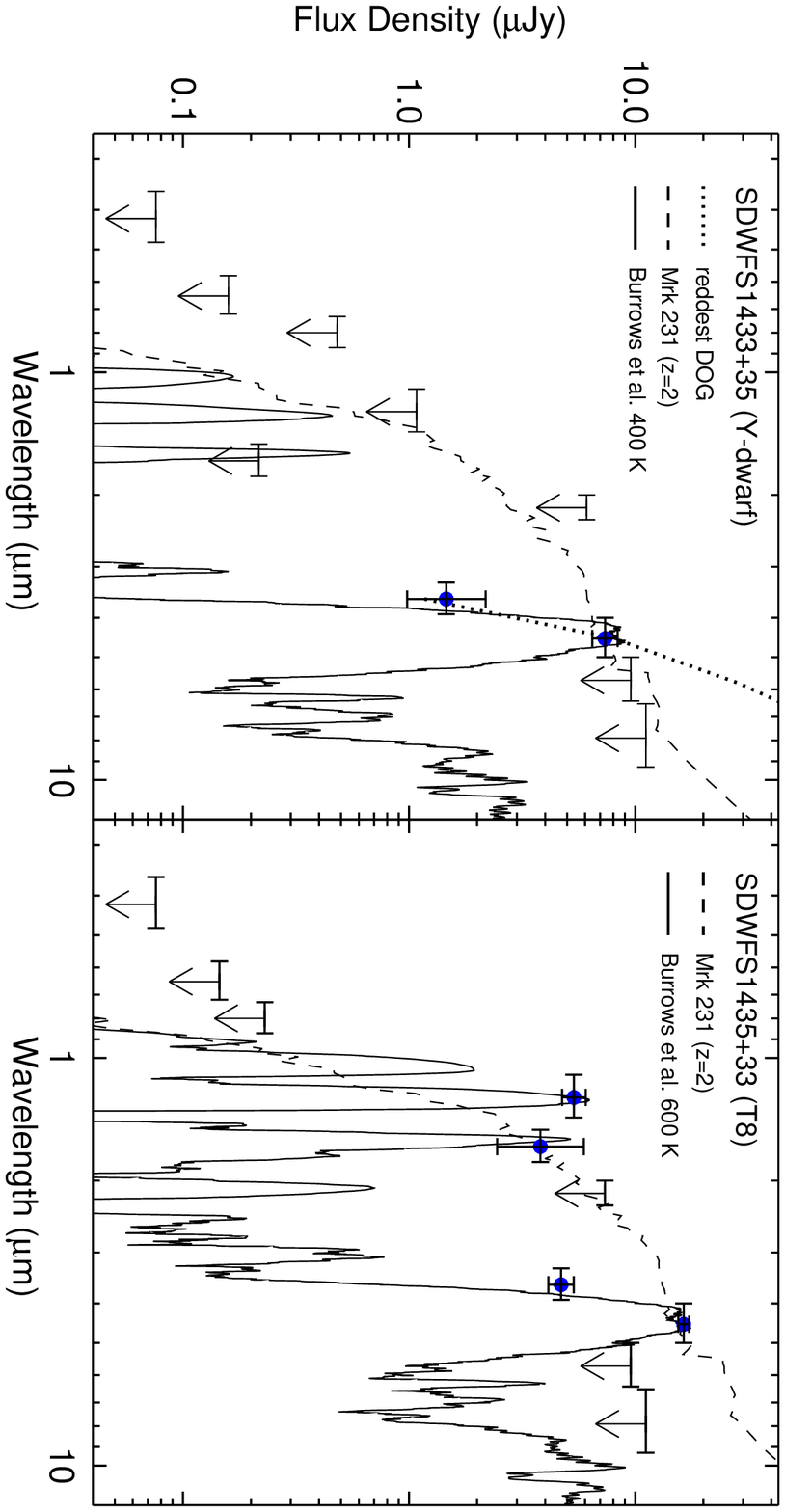}{5.5in}{90}{70}{70}{280}{-50}
\caption{SED's for the Y-dwarf (left) and T8 (right) candidates.  
The solid line shows 400K (left) and 600K (right) brown dwarf models
from \citet{Burrows:03}, normalized at $4.5\mu$m. 
The models assume non-equilibrium chemistry and $log(g) = 5.0$.  
The dashed line shows the spectrum of Mrk 231 at $z=2$, normalized at $4.5\mu$m, 
from \citet{Bussmann:09}, representative of DOG SED's.  
The dotted line (left) shows the SED for the reddest 
(in $[3.6] - [4.5]$) DOG found in our search, 
SDWFS~J143819.58+340957.3, which continues to rise steeply beyond $4.5\mu$m, but
is undetected below  $3.6\mu$m.  
\label{fig:BD_sed}}
\end{figure}

\begin{figure}
\plotfiddle{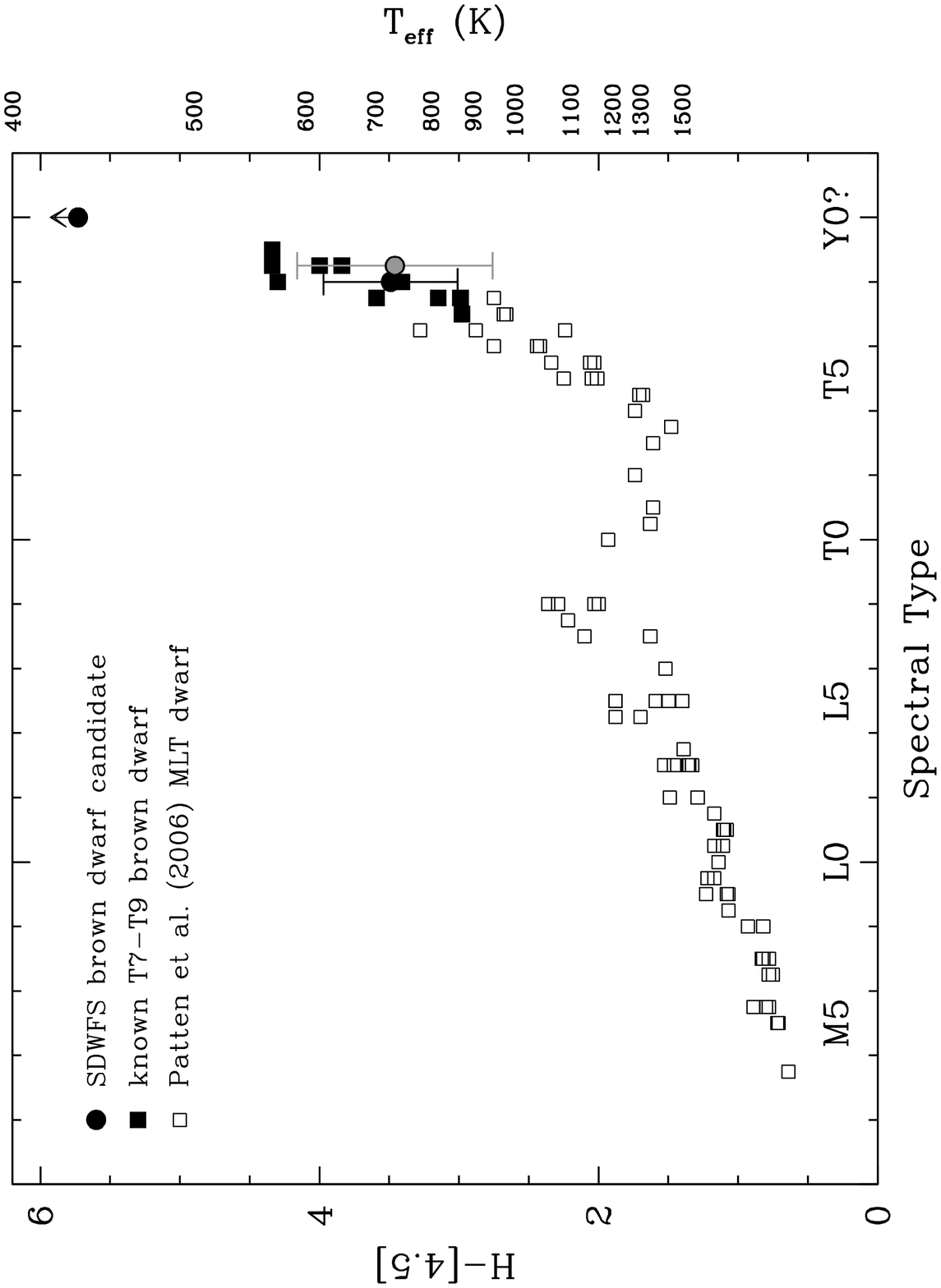}{5.5in}{-90}{70}{70}{-280}{420}
\caption{$H - [4.5]$ color vs. spectral type for brown dwarfs from \citet{Patten:06} (open squares).  
Filled squares show data for the published brown dwarfs 
with $[3.6] - [4.5] \geq 1.5$ from \citet{Patten:06}, \citet{Warren:07},
\citet{Burningham:08}, \citet{Burgasser:08} and \citet{Leggett:10}.
Filled circles show proposed spectral types and colors 
for three cool brown dwarfs from this SDWFS study: the observed 
$H - [4.5]$ for \teight, the lower limit (marked by an arrow) on $H - [4.5]$ for \ydwarf,
and the inferred $H - [4.5]$ for \teightnew\ (shown with grey shading) based on its measured
$J$ and $[4.5]$ photometry and an assumed $J-H = -0.35$.  
The error bar for \teightshortnew\ was increased to match the limit on
$H - [4.5]$ given in Table~1.  Model temperatures as in Fig.~1 
corresponding to $H - [4.5]$ are plotted on the right axis.
\label{fig:hm4p5}}
\end{figure}

However \citet[]{Patten:06} and \citet{Golimowski:04} show that 
the observed [4.5] or $M$-band flux is substantially lower 
than predicted by these models.  This is also apparent in Figure \ref{fig:colcol}. 
Detailed modeling for the known $C \geq 1.5$ brown dwarfs 
\citep[e.g. Figure 7 of ][]{Leggett:09} finds $T_{\rm eff} = 550 - 800$K
for these objects. The Burrows models for these temperatures
predict $[3.6] - [4.5]$ colors which are significantly redder than observed, 
and $[4.5] - [8]$ colors which are significantly bluer than observed,  
and suppressing the model [4.5] flux corrects this.

\begin{figure}
\plotone{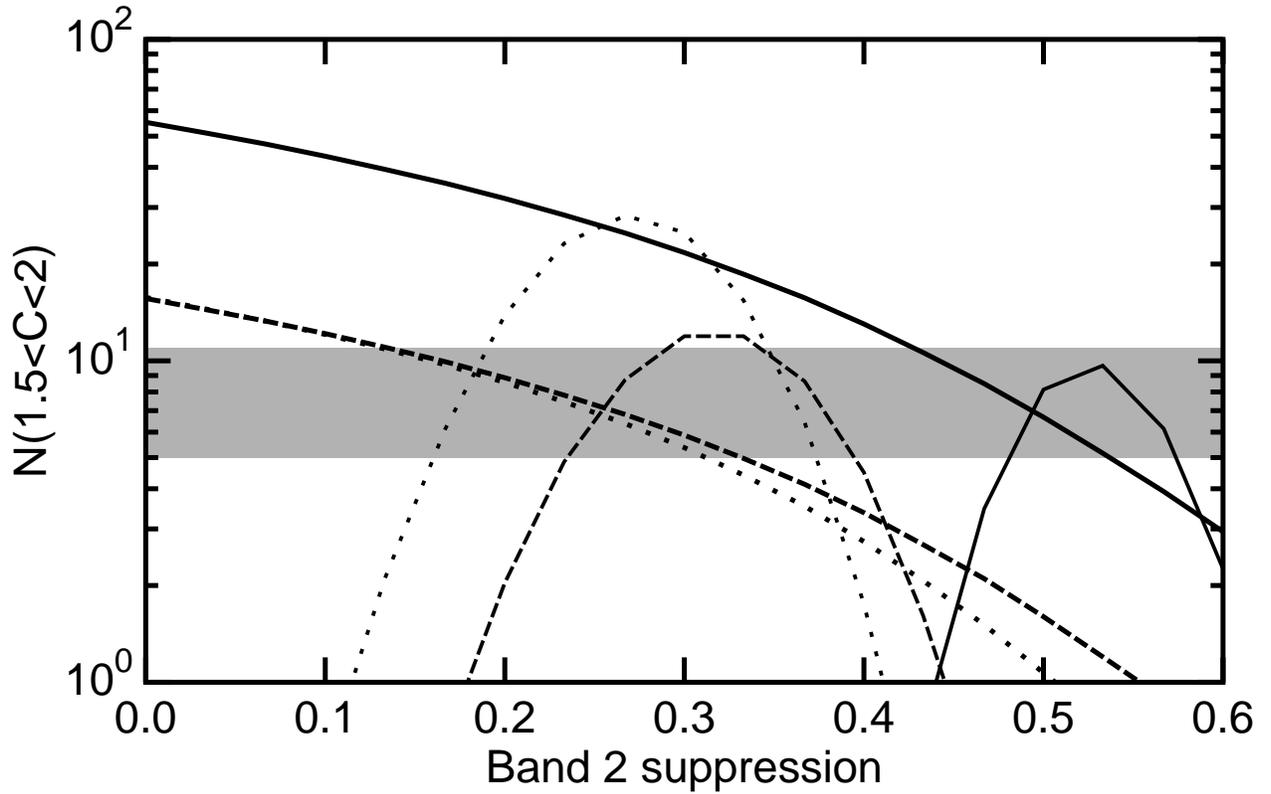}
\caption{The expected number (heavier lines) of SDWFS brown dwarfs
in the color range $1.5 < C < 2$, where $C \equiv [3.6] - [4.5]$, 
and the relative Poisson likelihood (lighter lines) based on the numbers seen with 
$1.5 < C < 2$ and with $C > 2$,
vs. [4.5] flux suppression, for the single object \citet[][solid]{Chabrier:03},
system \citet[][dashed]{Chabrier:03}
and \citet[][dotted]{Bochanski:09}
mass functions.  The horizontal band shows the estimated observed number
after correcting for contamination.     
\label{fig:number-vs-supression}}
\end{figure}

Suppressing the model [4.5] flux has a strong effect on both the ratio problem
and the number problem.  This flux suppression, presumably due to
the CO fundamental at $4.7\mu$m, has been attributed to non-equilibrium
chemistry altering the expected CO absorption depths \citep{Hubeny:07}.
Suppression of flux in the spectrum causes some backwarming, 
so the effective temperature increases to
\begin{equation}
T_{\rm eff}^\prime = T_{\rm eff}/(1-Sf_2/R_2)^{1/4}
\end{equation} 
where $f_2$ is the fraction of the bolometric luminosity 
in the IRAC channel 2 ([4.5]) detection band from the model, 
$S$ is the [4.5] suppression, and $R_2$ is the resolution of
the IRAC $4.5 \mu$m filter.  We use $R_2 = 3$ which is somewhat lower
than the actual $R_2 = 4.5$, to allow for suppression of flux
outside of the [4.5] passband.  The $4.5 \mu$m
flux fraction decreases to $f_2^\prime = (1-S)f_2(T_{\rm eff}^\prime)$.
Increased suppression decreases the ratio $N(C > 2)/N(1.5 < C < 2)$
and also reduces the expected number counts, as shown by the heavy solid
curve in Figure \ref{fig:number-vs-supression}.  The horizontal
band shows the estimated range of 5 to 11 with $1.5 < C <2$.  The
lighter solid curve peaking near $S = 0.52$ shows the Poisson likelihood 
of a given flux suppression based on the numbers seen
in the two color bins, assuming that the \citet{Chabrier:03} single object 
mass function and uniform age distribution are correct.  The
likelihood is maximized by a suppression of $S = 0.52$. 
Similar results are obtained for a power law mass function
with $\alpha = 1.3$.  
This suppression of $S = 0.52$ agrees well with the estimate 
by \citet{Golimowski:04} that the [4.5] flux is suppressed by 
a factor between 1.5 and 2.5, which corresponds to $S = 0.33$ to 0.60
in our terminology.
In other words, suppressing the \citet{Burrows:03} model by a factor of two
brings both the models into agreement with both the 
observed mid-infrared colors and number counts.    

Another potential solution is to adopt a different mass function, which, 
like the predicted luminosities and colors, 
is not well known for these very low mass objects.  
Since the \citet{Burrows:03} models give the 
luminosity vs. mass and age for single brown dwarfs, 
it is not strictly correct to use system mass functions in
these calculations, but to provide a range of examples 
we have included  in Figure 6 both the \citet{Chabrier:03} single
object and system mass function, as well as the \citet{Bochanski:09} 
mass function. 
Based on SDSS observations of late M dwarfs, \citet{Bochanski:09}
find a log normal mass function peaking at $0.27 M_\odot$.  
This results in a much smaller predicted number density of brown dwarfs than for the
\citet{Chabrier:03} single object mass function, and requires a smaller 4.5 $\mu$m
flux suppression ($S =0.26$) to agree with the observed SDWFS
counts, as shown by the dotted curves in Figure \ref{fig:number-vs-supression}.  
However, this would not account as well for the observed 
$4.5\mu$m flux discrepancy. 
The \citet{Chabrier:03} system mass function, which peaks at
$0.2 M_\odot$, shows intermediate results, matching the SDWFS counts for
a flux suppression of $S = 0.31$, as shown by the dashed curves 
in Figure \ref{fig:number-vs-supression}. 
The number density data shown in \citet{Bochanski:09} 
flatten significantly at the low mass end, 
so this log normal mass function may not be reliable in the brown dwarf regime.
For both the \citet{Bochanski:09} mass function and
significant flux suppression of the Burrows models to be correct  
would suggest that nearly all of the SDWFS brown dwarf candidates are dusty
galaxies with no evidence for star formation in the rest UV or mid-IR.

Scaling from the SDWFS counts to the all-sky {\it WISE} survey,
which launched  on 2009 Dec. 14, is simpler than comparing to
models.  The {\it WISE} sensitivity requirement at $4.6\mu$m is $160\mu$Jy,
resulting in a surveyed volume which is $\sim40 \times$ greater
than SDWFS, and hence $\sim 250$ to 500 similarly cool brown dwarfs for a
Euclidean distribution.  
These {\it WISE} brown dwarfs will be the nearest examples, 
with correspondingly brighter fluxes and larger parallaxes and proper motions, 
making followup observations much easier.  
This should enable a definitive determination
of the properties of the ultracool brown dwarf population.

\acknowledgements

The authors thank 
Emanuele Daddi, Mark Dickinson, Jason Melbourne, and Tom Soifer for assistance obtaining observations, 
Nick Seymour for assistance with SDWFS and WIRC reductions, and 
Vandana Desai for the Mrk 231 spectrum and information about DOG SED's.   
Tom Soifer, Marcia Rieke, Dan Weedman, and Jim Houck are thanked for allowing access to the GTO MIPS survey of the NDWFS, and we acknowledge Buell Jannuzi's central role in
the NDWFS and related surveys of the field.  
Discussions with Roc Cutri helped us understand the mid-IR characteristics of AGB stars,
and Szymon Kozlowski clarified questions about SDWFS variability measurements. 
We thank the anonymous referee for a detailed and careful 
review which improved the accuracy of the presentation.  
This work is based on observations made with the {\it Spitzer Space
Telescope}, which is operated by the Jet Propulsion Laboratory,
California Institute of Technology under contract with NASA.  
This work made use of images and data products provided by the 
NOAO Deep Wide-Field Survey (NDWFS), which is supported by the 
National Optical Astronomy Observatory (NOAO), and followup NOAO surveys. 
NOAO is operated by AURA, Inc., 
under a cooperative agreement with the National Science Foundation.
Some of the data presented herein were obtained at the W.M. Keck Observatory, 
which is operated as a scientific partnership among  
Caltech, the University of California and NASA. 
The Keck Observatory was made possible by the generous financial support 
of the W.M. Keck Foundation, which also provided support for MB.  
Some data was obtained at the Hale Telescope, 
Palomar Observatory as part of a continuing collaboration 
between Caltech, NASA/JPL, and Cornell University. 
Support for this work was provided by NASA through an award issued by
JPL/Caltech. 
AHG acknowledges support for this work by the National Science 
Foundation (NSF) under Grant No. 0708490.  
Support for AMG and QK's contribution to this work was provided by  
the NSF Science \& Technology Center for AO, managed by UCSC  
(AST-9876783), and the Levine-Leichtman Family Foundation.

Facilities:  
\facility{{\it Spitzer Space Telescope} (IRAC) (MIPS), Palomar 200''
(WIRC), Keck (NIRC2), KPNO (NEWFIRM)}


\bibliographystyle{apj}
\bibliography{apj-jour,thebibliography}

 
%
 

\begin{deluxetable}{lcccccccccccl}
\tabletypesize{\tiny}
\tablewidth{0pt}
\tablecaption{SDWFS Ultracool Brown Dwarf Candidates.}
\rotate
\tablehead{
\colhead{ID} & 
\colhead{$B_W$} & \colhead{$R$} & \colhead{$I$} &
\colhead{$J$} & \colhead{$H$} & \colhead{$K_s$} &
\colhead{[3.6]} & \colhead{[4.5]} & \colhead{[5.8]} & \colhead{[8.0]} &
\colhead{$[3.6] - [4.5]$} & \colhead{Notes}}
\startdata
SDWFS J142822.12+331056.5 & $>25.9$ & $>24.4$ & $>22.8$ & $>22.0$ & $>20.6$ & $>19.7$ & $20.03\pm0.24$ & $18.27\pm0.11$ & $>17.7$ & $>16.9$ & $1.76\pm0.26$ &  \\
SDWFS J142831.46+354923.0 & $>26.5$ & $>25.0$ & $>24.0$ & $>21.9$ & $>21.4$ & $>19.0$ & $19.20\pm0.11$ & $17.66\pm0.07$ & $>17.7$ & $16.39\pm0.36$ & $1.54\pm0.13$ & a \\
SDWFS J143222.82+323746.5 & $>26.7$ & $>25.6$ & $>24.5$ & $21.17\pm0.18$ & $>20.8$ & $>20.2$ & $19.97\pm0.22$ & $18.06\pm0.09$ & $>17.7$ & $>16.9$ & $1.91\pm0.24$ &  \\
SDWFS J143355.24+343422.7 & $>26.7$ & $>25.2$ & $>24.0$ & $>21.7$ & $>21.3$ & $>19.8$ & $20.30\pm0.30$ & $18.48\pm0.14$ & $>17.7$ & $>16.9$ & $1.82\pm0.33$ & b \\
SDWFS J143356.62+351849.2 & $>26.7$ & $>25.7$ & $>24.2$ & $>22.9$ & $>24.2$ & $>20.1$ & $20.71\pm0.44$ & $18.47\pm0.14$ & $>17.7$ & $>16.9$ & $2.24\pm0.46$ & c \\
SDWFS J143524.44+335334.6 & $>26.7$ & $>25.8$ & $>25.0$ & $21.16\pm0.13$ & $21.09\pm0.48$ & $>19.9$ & $19.44\pm0.14$ & $17.60\pm0.06$ & $>17.7$ & $>16.9$ & $1.84\pm0.15$ & d \\
SDWFS J143531.65+344509.4 & $>26.8$ & $>24.8$ & $>24.0$ & $>21.8$ & $>20.5$ & $>20.3$ & $19.94\pm0.22$ & $18.28\pm0.11$ & $>17.7$ & $>16.9$ & $1.66\pm0.25$ &  \\
SDWFS J143555.04+344307.0 & $>26.3$ & $>25.2$ & $>23.1$ & $>21.8$ & $>20.5$ & $>19.5$ & $19.94\pm0.22$ & $18.38\pm0.12$ & $>17.7$ & $>16.9$ & $1.56\pm0.25$ &  \\
SDWFS J143605.72+342834.5 & $>26.6$ & $>25.3$ & $>24.6$ & $>21.9$ & $>20.9$ & $>20.0$ & $20.10\pm0.25$ & $18.49\pm0.14$ & $>17.7$ & $>16.9$ & $1.61\pm0.29$ &  \\
SDWFS J143712.48+334516.5 & $>25.1$ & $>23.4$ & $>21.9$ & $>21.2$ & $>21.3$ & $>20.3$ & $19.93\pm0.22$ & $18.32\pm0.12$ & $16.80\pm0.25$ & $>16.9$ & $1.61\pm0.25$ & e \\
SDWFS J143724.88+343950.9 & $>26.8$ & $>25.3$ & $>24.1$ & $>21.9$ & $>21.4$ & $>20.3$ & $19.94\pm0.22$ & $18.43\pm0.13$ & $>17.7$ & $>16.9$ & $1.51\pm0.26$ & f \\
SDWFS J143749.23+333657.7 & $>26.8$ & $>25.8$ & $>24.4$ & $>21.8$ & $>20.5$ & $>20.7$ & $19.84\pm0.20$ & $18.28\pm0.11$ & $>17.7$ & $16.31\pm0.33$ & $1.56\pm0.23$ & g \\
SDWFS J143819.26+334856.5 & $>26.4$ & $>25.3$ & $>24.5$ & $>22.1$ & $>20.6$ & $>20.2$ & $19.91\pm0.21$ & $18.38\pm0.13$ & $17.45\pm0.45$ & $>16.9$ & $1.53\pm0.25$ &  \\
SDWFS J143821.36+353523.3 & $>27.0$ & $>24.8$ & $>24.3$ & $>22.4$ & $>21.3$ & $>20.1$ & $19.93\pm0.22$ & $18.38\pm0.12$ & $>17.7$ & $>16.9$ & $1.55\pm0.25$ & h \\
\hline
\multicolumn{13}{c}{Likely AGN} \\
\hline
SDWFS J142506.42+350526.0 & $25.38\pm0.22$ & $24.43\pm0.32$ & $>24.4$ & $>21.5$ & $>21.3$ & $>19.3$ & $20.23\pm0.28$ & $18.46\pm0.13$ & $16.96\pm0.29$ & $16.53\pm0.41$ & $1.77\pm0.31$ & \\
SDWFS J143334.06+344009.3 & $25.68\pm0.31$ & $>25.3$ & $>24.3$ & $>21.7$ & $>21.5$ & $>19.8$ & $19.88\pm0.21$ & $18.36\pm0.12$ & $17.41\pm0.44$ & $>16.9$ & $1.52\pm0.24$ & \\
SDWFS J143359.13+331454.8 & $25.95\pm0.41$ & $>25.8$ & $>24.6$ & $>21.8$ & $>21.6$ & $>19.8$ & $19.76\pm0.18$ & $18.26\pm0.11$ & $>17.7$ & $>16.9$ & $1.50\pm0.21$ & \\
SDWFS J143833.76+352209.2 & $26.55\pm0.45$ & $>25.8$ & $>25.0$ & $>21.7$ & $>21.0$ & $>20.5$ & $20.30\pm0.30$ & $18.47\pm0.14$ & $>17.7$ & $>16.9$ & $1.83\pm0.33$ &  \\
\enddata

\tablecomments{Photometry is all Vega-based, total magnitudes.
Photometry is from NDWFS ($B_WRI$; Jannuzi et al., in prep.), 
and NEWFIRM ($JHK_s$; Gonzalez et al., in prep.).
IRAC photometry is from SDWFS (Ashby et al. 2009).
Non-detection limits are the 2$\sigma$ limits for the relevant bands 
(see \S \ref{sec:select} and \ref{sec:OtherData} for details). 
a: $J$ and $H$ from Palomar (\S2.3), $K_s$ from NDWFS; 
b: marginal J detection;
c: Y-dwarf candidate, $H$ from Keck (\S2.3); 
d: T8 candidate; 
e: Bleed trail makes source questionable; 
f: [4.5] morphology makes source questionable; 
g: [8] detection appears spurious;  
h: Possibly variable.}

\label{table}
\end{deluxetable}
\normalsize

\clearpage

\end{document}